\documentclass[manuscript, screen]{acmart}

\AtBeginDocument{%
  }

\setcopyright{none}
\copyrightyear{2027}
\acmYear{2027}
\acmDOI{}
\acmConference[CHI '27]{Proceedings of the 2027 CHI Conference on Human Factors in Computing Systems}{May 10--14, 2027}{Pittsburgh, PA, USA}
\acmISBN{}
\usepackage{algorithm}
\usepackage{algorithmic}
\usepackage{amsmath}
\usepackage{booktabs}
\usepackage{array}
\usepackage{xcolor}
\usepackage{colortbl}
\newcommand{\runner}[1]{\underline{#1}}

\begin{document}

\title{Catellect-VL-2B: A Vision-Language Model for Edge-Based Feline Behavior Understanding}

\author{YuHang Wu}
\affiliation{%
  \institution{Catellect Research}
  \country{USA}
}

\author{JieXian Liu}
\affiliation{%
  \institution{Catellect Research}
  \country{USA}
}

\author{Junyi Wang}
\affiliation{%
  \institution{Catellect Research}
  \country{USA}
}
\author{Jia Tao}
\authornote{Corresponding author.}
\affiliation{%
  \institution{Catellect Research}
  \country{USA}
}
\email{hello@catellect.com}

\renewcommand{\shortauthors}{Wu et al.}

\begin{abstract}
The task of Feline Behavior Understanding requires models that can identify subtle visual cues, keep behavior interpretations auditable, and support low-latency, privacy-sensitive deployment. Directly prompting general Vision-Language Models (VLMs) is poorly suited to this setting: instead of first reporting visible evidence such as ear position and tail posture, they may jump directly to labels such as relaxed, afraid, or in pain. This makes the output difficult to verify and poorly aligned with edge-based use, where compact JSON outputs are preferable to long free-form explanations. We present Catellect-VL-2B, an edge-based VLM for Feline Behavior Understanding that generates JSON-formatted Structured Output for feline behavior. Catellect-VL-2B is post-trained from Qwen3-VL-2B on CatellectBench, our 40K-sample image--behavior annotation dataset with approximately 38K stage-specific training instances and a 2K held-out test set. The multi-phase Post-Training recipe combines natural-language behavior warmup, Field-Aware Weighted(FAW) supervised fine-tuning, and compact behavior serialization. Experimental results show that Catellect\mbox{-}VL\mbox{-}2B equipped with compact output serialization achieves 80.62\% behavior\mbox{-}field macro accuracy, and delivers a 2.51$\times$ speedup over its full\mbox{-}JSON baseline of identical parameter size when deployed on the RK3576 edge chip, making it suitable for edge deployment. We further build a 3K-entry feline behavior knowledge base that maps structured behavior fields to emotion and intent concepts for evidence-grounded interpretation. Together, these results show that Catellect-VL-2B can make animal-centered VLMs more accurate, auditable, and deployable. 

\end{abstract}



\maketitle

\section{Introduction}

\begin{figure*}[t]
\centering
\includegraphics[width=1.0\textwidth]{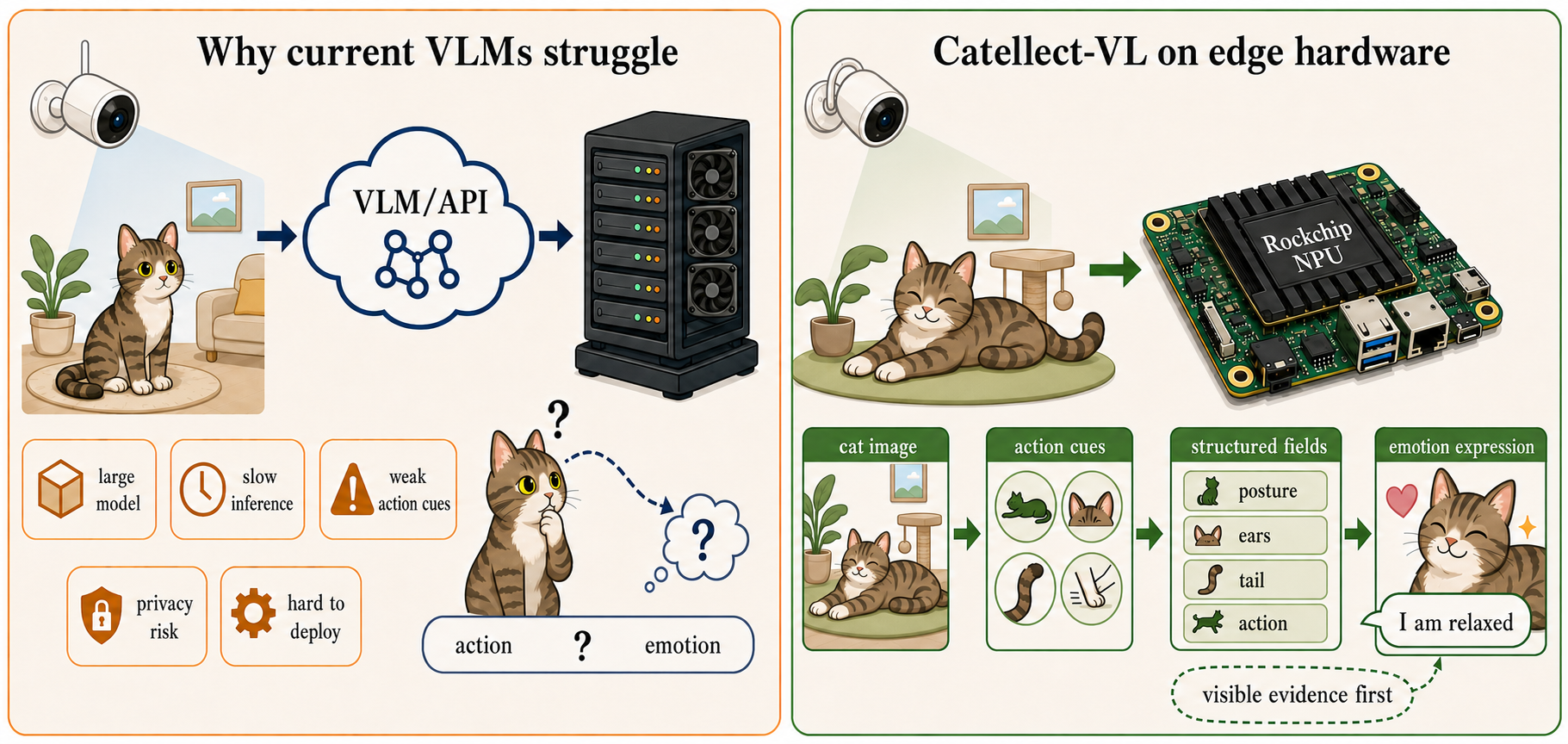}
\caption{Motivation for structured Feline Behavior Understanding. General VLM and API-based solutions struggle to provide accurate, private, and low-latency behavior understanding for practical deployment. Catellect-VL-2B instead maps cat images to structured behavior fields, which are then matched against a feline behavior knowledge base for evidence-supported interpretation.}
\Description{A workflow diagram contrasting general vision-language model prompting with the Catellect-VL-2B structured behavior prediction and knowledge-grounding pipeline.}
\label{fig:intro}
\end{figure*}

Recent vision-language models (VLMs), including Qwen3-VL~\cite{qwen2025qwen3vl}, GPT~\cite{openai2025gpt5systemcard}, MiniCPM-V 4.5~\cite{yu2026minicpmv45}, and Claude~\cite{anthropic2025claude4systemcard}, have substantially advanced general-purpose visual understanding. For companion-animal care, however, recognizing an animal is only the first step~\cite{hu2026meowomni1multimodallarge}. A more consequential question is: \textbf{Can a model help people understand what a cat is doing and identify the visible cues that support that interpretation?} Cats communicate comfort, fear, attention, pain, and practical needs through combinations of body posture, ear and tail position, actions, and interactions with nearby objects~\cite{ellis2018felineemotions,evangelista2019feline}. These signals can be subtle, transient, and visually ambiguous. A generic caption such as ``a cat lying on the floor'' may be correct while omitting the cues that matter for interpretation. Feline Behavior Understanding is therefore a fine-grained, structured visual understanding task with direct implications for how people monitor and respond to companion animals, as illustrated in Figure~\ref{fig:intro}.

To address this task, we introduce Catellect-VL-2B, a 2B-parameter VLM designed for edge-based Feline Behavior Understanding. Catellect-VL-2B is post-trained from Qwen3-VL-2B on CatellectBench, our 40K-sample image--behavior annotation dataset with approximately 38K stage-specific training instances and a 2K held-out test set. Given a cat image, the model generates JSON-formatted Structured Output that records observable behavior fields rather than directly producing an unconstrained emotional judgment. Catellect-VL-2B is obtained through the following three-phase Post-Training recipe:

\noindent\textbf{1) Natural-Language Behavior Warmup.} The first phase transfers feline behavior semantics from teacher-generated natural-language descriptions to the 2B model, establishing domain knowledge before structured prediction is introduced.

\noindent\textbf{2) Field-Aware Weighted Supervised Fine-Tuning.} The second phase maps each image to a fixed JSON schema covering cat presence, action, body posture, ears, tail, face, fur, and environmental interactions. A field-weighted alignment (FAW) loss emphasizes behavior-bearing values over JSON syntax, making the Structured Output both parseable and behavior focused.

\noindent\textbf{3) Compact Behavior Serialization.} The third phase converts the same behavior schema into a shorter representation that can be deterministically restored to JSON. This reduces token-by-token generation cost without removing visual tokens or requiring a server-oriented inference framework. After Phase 3, Catellect-VL-2B uses compact output serialization and is deployed for local inference on a Rockchip RK3576 platform with 8~GB of memory~\cite{rockchip2024rk3576}.

For downstream interpretation, we construct a 3K-entry feline behavior knowledge base that maps predicted behavior fields to emotion concepts while retaining the visual evidence supporting each interpretation. Details of CatellectBench and the knowledge base are provided in Section~\ref{sec:bench}.

Experimental results show that the Phase 2 JSON model achieves 83.63\% behavior\mbox{-}field macro accuracy, outperforming Qwen3.5\mbox{-}VL\mbox{-}9B by 8.55\% while using about one\mbox{-}fifth of its parameters.
After Phase 3, Catellect\mbox{-}VL\mbox{-}2B with compact output serialization achieves 80.62\% macro accuracy and delivers a $2.98\times$ speedup over the same\mbox{-}size model baseline under our Nvidia A100 diagnostic setting, retaining approximately 96.4\% of the Phase 2 accuracy.
When deployed on the RK3576 edge chip, it further yields a $2.51\times$ speedup relative to its long\mbox{-}JSON baseline. The knowledge base further reaches 100.0\% top\mbox{-}1 retrieval accuracy across behavior categories.
These results show that domain\mbox{-}specific post\mbox{-}training and compact structured output can jointly support accurate, auditable, and edge\mbox{-}deployable feline behavior understanding. Our contributions are:

\begin{itemize}
    \item This paper formulates Feline Behavior Understanding as a structured vision-language task and builds CatellectBench, a 40K-sample image--behavior annotation dataset for Post-Training and evaluation.
    \item This paper develops Catellect-VL-2B using a three-phase Post-Training recipe that combines natural-language behavior warmup, Field-Aware Weighted supervised fine-tuning with FAW loss, and compact behavior serialization to learn structured behavior fields while enabling faster edge-based inference.
    \item This paper constructs a 3,151-entry feline behavior knowledge base organized around 120 behavior concepts for mapping structured behavior fields to feline emotion and intent concepts through exact field-rule matching, thereby producing evidence-supported interpretations.
\end{itemize}
\begin{table*}[t]
\centering
\scriptsize
\renewcommand{\arraystretch}{1.10}
\setlength{\tabcolsep}{3pt}
\begin{tabular*}{\textwidth}{@{\extracolsep{\fill}}p{0.25\textwidth}p{0.45\textwidth}p{0.25\textwidth}@{}}
\toprule
\textbf{Schema Field} & \textbf{Label Candidates} & \textbf{Explanation} \\
\midrule
\texttt{cats\_visible} & integer count, where 0 means no visible cat & Cat presence and count evaluation \\
\texttt{lighting}, \texttt{other\_beings[]} & lighting enum; being id, type, short description & Image-level context and negative cases \\
\texttt{cats[].location\_on} & floor, bed, sofa, chair, table, cabinet, window, cat tree, cat bed, human lap, suspended, other, unknown & Ground the cat in household space \\
\texttt{cats[].vertical\_position} & ground, low, middle, high, suspended, unknown & Distinguish floor-level and elevated behavior \\
\texttt{cats[].nearby\_anchors[]} & anchor type, visual id, proximity & Link behavior to objects such as food bowls, water bowls, litter boxes, doors, and windows \\
\texttt{cats[].action} & sleep, still, active, jump, eating, drinking, grooming, toileting, vomiting, other, unknown & Main behavior field \\
\texttt{cats[].posture.overall\_body} & crouching, side-lying, prone, standing, arched, raised-hip, curled, stretched, half-rising, suspended, other, unknown & Whole-body posture cue \\
\texttt{cats[].posture.ears.position} & upright, forward, side-flattened, fully backward, unilateral abnormal, unknown & Ear cue for attention or stress \\
\texttt{cats[].posture.tail.position} & upright, question-mark, slightly curved, horizontal, down, tucked, puffed, curled, unknown & Tail cue for affect and arousal \\
\texttt{cats[].posture.face} & face visible; eyelid enum; mouth enum & Facial cue and hallucination control \\
\texttt{cats[].posture.fur\_state} & normal or piloerection & Arousal and risk-related body cue \\
\texttt{abnormalities[]}, \texttt{environment\_anomalies[]} & tag, short visual description, severity hint & Health-related or environment risk cues \\
\texttt{interactions[]} & participants and interaction type & Ground social and object interaction \\
\bottomrule
\end{tabular*}
\caption{Structure of the JSON ground-truth annotations in CatellectBench. Each row lists a schema field, its label space, and its role in feline behavior.}
\label{tab:schema}
\end{table*}

\section{Related Work}

\subsection{Traditional Animal Behavior Interpretation}
The computational study of animal behavior has progressed from manual ethogram coding to automated perception pipelines. DeepLabCut \citep{mathis2018deeplabcut} and SLEAP \citep{pereira2022sleap} estimate animal poses, DeepEthogram \citep{bohnslav2021deepeethogram} recognizes behavior from video, and SuperAnimal \citep{ye2024superanimal} supports pose estimation across species. Feline-specific research further demonstrates the importance of fine-grained visual evidence: the Feline Grimace Scale defines facial action units for pain assessment \citep{evangelista2019feline}, while CatFLW provides facial landmarks for automated analysis \citep{martvel2023catflw}. These methods establish strong foundations for animal perception, but their outputs are typically pose tracks, landmark coordinates, or discrete behavior labels rather than a unified, language-compatible record of multiple observable feline cues.

\subsection{The Rise of Vision-Language Models}
Recent VLMs connect visual encoders with large language models and provide a general interface for instruction following and open-ended visual reasoning \citep{alayrac2022flamingo,li2023blip2,dai2023instructblip,zhu2023minigpt4,liu2023visual,bai2023qwenvl,chen2024internvl,qwen2025qwen3vl}. These models are primarily developed for general-purpose, human-centered tasks, whereas animal behavior understanding depends on non-verbal and species-specific signals. Model scale creates an additional deployment gap: large general VLMs can exceed edge memory and compute budgets, while reducing parameter count alone does not resolve domain-specific accuracy or autoregressive generation latency. Meow-Omni-1 \citep{hu2026meowomni1multimodallarge} moves toward animal-centered foundation models by integrating video, audio, physiological time series, and text for feline intent inference. However, image-based VLMs that jointly predict multiple observable feline behavior fields remain underexplored. Catellect-VL-2B focuses on this setting by adapting a 2B-parameter VLM to produce a fixed, field-level behavior representation from a single image.


\section{CatellectBench and Feline Behavior Knowledge}
\label{sec:bench}
\subsection{Preliminaries}

\noindent\textbf{Veterinarian-Designed Schema.} Feline behavior understanding maps an image $x$ to a structured behavior report $y$ under a fixed schema $\mathcal{S}$. Throughout this paper, structured behavior report denotes a parseable JSON object that follows $\mathcal{S}$, while structured behavior fields denote the normalized values contained in that report. A PhD-trained veterinarian with a valid veterinary practice license designed the schema and manually annotated a 416-image gold seed set. Table~\ref{tab:schema} summarizes the main schema fields.

\begin{figure*}[t]
\centering
\includegraphics[width=0.9\textwidth]{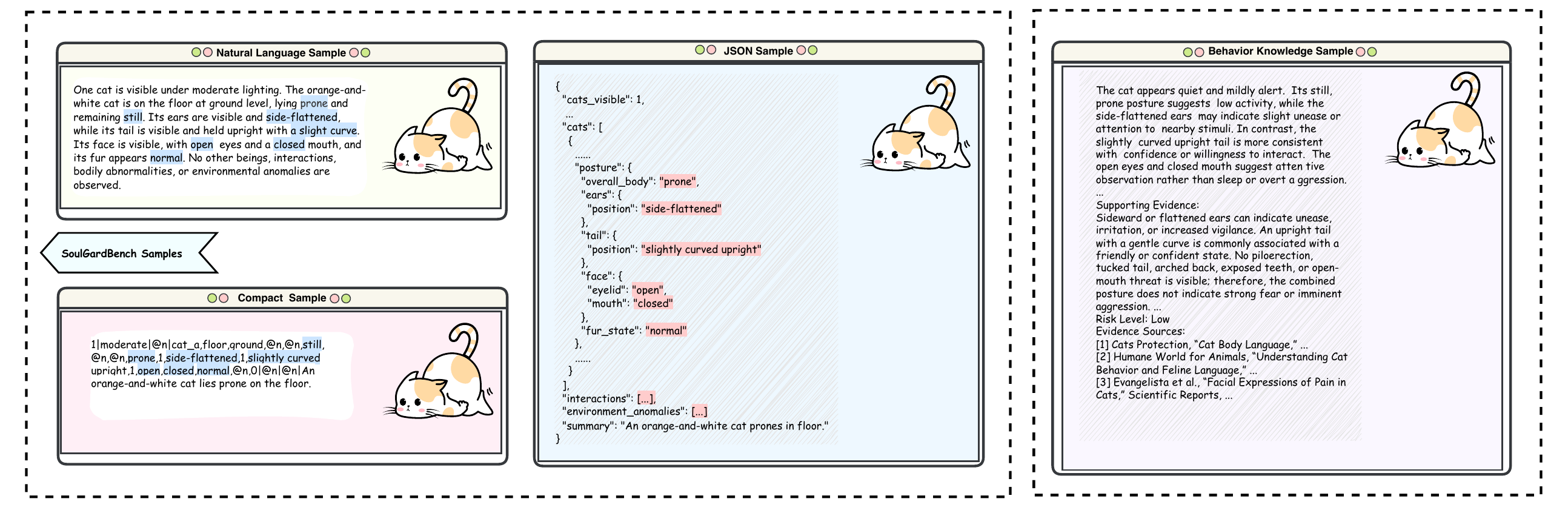}
\caption{Examples of the three CatellectBench training targets and the downstream knowledge-grounded interpretation. For the same image, Phases 1--3 use a controlled natural-language description, a structured behavior report in JSON format, and a compact serialization, respectively. The held-out test set uses JSON ground truth. The rightmost panel shows how predicted JSON fields are matched against the feline behavior knowledge base to obtain an evidence-supported interpretation. Training details are provided in Section~\ref{sec:post-training}.}
\label{fig:target_formats}
\end{figure*}

\noindent\textbf{Veterinary Basis.} The schema follows how feline behavior is assessed in veterinary practice. Prior study emphasizes jointly observing action, body posture, ears, tail, face, and fur state when assessing feline emotional motivation \citep{ellis2018felineemotions}. AAFP/ISFM guidelines further identify feeding, drinking, toileting, resting, scratching, play areas, safe places, and elevated locations as important aspects of feline welfare and behavior \citep{ellis2013aafpisfm}. We also retain explicit face fields because facial cues support both feline pain assessment and automated facial analysis \citep{evangelista2019feline,martvel2023catflw}. Accordingly, the schema treats action, posture, body-part cues, environment, and interaction as explicit fields rather than incidental caption tokens.

\noindent\textbf{Teacher Model Selection for Scaled Data.} To scale the veterinarian-designed annotations beyond the 416-image gold seed set, we evaluate three candidate teacher models for large-scale annotation: Doubao-Seed-2.0-Pro, Gemini3-Pro-Preview, and GPT-5.4. Each candidate annotates the same 416 images under schema $\mathcal{S}$, and we compute field-wise exact-match agreement with the veterinarian annotations. A field is counted as correct only when its canonical value exactly matches the veterinarian value at the corresponding schema path; no semantic relaxation or language-model judging is used. As shown in Table~\ref{tab:teacher_selection}, Doubao-Seed-2.0-Pro achieves the highest agreement and is therefore selected to annotate the filtered image pool.

\begin{table}[t]
\centering
\small
\setlength{\tabcolsep}{0pt}
\begin{tabular*}{\columnwidth}{@{\extracolsep{\fill}}lrr@{}}
\toprule
\textbf{Teacher candidate} & \textbf{Agreement (\%)} $\uparrow$ & \textbf{Cost (USD)} $\downarrow$ \\
\midrule
Doubao-Seed-2.0-Pro & \textbf{89.18} & \textbf{2.8} \\
GPT-5.4 & 82.78 & 17.5 \\
Gemini3-Pro-Preview & 75.19 & 14.0 \\
\bottomrule
\end{tabular*}
\caption{Teacher-model selection on the 416-image veterinarian-annotated gold seed set.}
\label{tab:teacher_selection}
\end{table}

\begin{algorithm}[t]
\caption{Distilled Feline Behavior Dataset Construction}
\label{alg:dataset}
\begin{algorithmic}[1]
\REQUIRE Raw images and video frames $\mathcal{X}_{\mathrm{raw}}$, CLIP threshold $\tau$, veterinarian-designed schema $\mathcal{S}$, veterinarian-labeled seed set $\mathcal{V}_{\mathrm{gold}}$, teacher candidates $\mathcal{T}$
\STATE Encode images with OpenAI CLIP ViT-B/32 and keep a filtered pool $\mathcal{X}$ by requiring maximum retained-set cosine similarity $<\tau$.
\STATE Evaluate each $T\in\mathcal{T}$ on $\mathcal{V}_{\mathrm{gold}}$ by field-wise exact-match agreement.
\STATE Select $T^{*}\leftarrow\arg\max_{T\in\mathcal{T}}\mathrm{Agree}(T,\mathcal{V}_{\mathrm{gold}})$.
\STATE $\mathcal{D}\leftarrow\varnothing$
\FOR{each image or sampled frame $x\in\mathcal{X}$}
    \STATE Generate a schema-constrained structured behavior report $\tilde{y}\leftarrow T^{*}(x,\mathcal{S})$.
    \IF{$x$ is readable and $\tilde{y}$ is valid JSON under $\mathcal{S}$}
        \STATE Canonicalize enum values, empty fields, and unknown fields.
        \STATE Add $(x,\tilde{y})$ to $\mathcal{D}$.
    \ENDIF
\ENDFOR
\STATE Remove missing images and duplicated reports from $\mathcal{D}$.
\STATE Split $\mathcal{D}$ into a structured training set $\mathcal{D}_{\mathrm{json}}$ and a held-out test set $\mathcal{D}_{\mathrm{test}}$ at a 10:1 ratio.
\STATE Derive the natural-language set $\mathcal{D}_{\mathrm{nat}}$ and compact set $\mathcal{D}_{\mathrm{cmp}}$ from selected samples in $\mathcal{D}_{\mathrm{json}}$.
\RETURN $\mathcal{D}_{\mathrm{nat}}$, $\mathcal{D}_{\mathrm{json}}$, $\mathcal{D}_{\mathrm{cmp}}$, $\mathcal{D}_{\mathrm{test}}$
\end{algorithmic}
\end{algorithm}

\subsection{Dataset Construction}

We build CatellectBench from self-collected household cat images and video frames. The raw pool contains more than 100K images or sampled frames covering six household cats in ordinary home environments. Because adjacent frames and burst photos can be visually near-duplicate, we first remove redundant images with an OpenAI CLIP ViT-B/32 image encoder \citep{radford2021learning}. For each image $x_i$, we compute an $\ell_2$-normalized CLIP image embedding
\begin{equation}
z_i=\frac{f_{\mathrm{CLIP}}(x_i)}{\|f_{\mathrm{CLIP}}(x_i)\|_2}.
\end{equation}
During filtering, a candidate image is kept only when its maximum cosine similarity to the retained set is below $\tau=0.8$:
\begin{equation}
\max_{x_j\in\mathcal{R}} z_i^\top z_j < \tau,
\end{equation}
where $\mathcal{R}$ is the current retained pool. This CLIP filtering step reduces repeated views while preserving behavior diversity.

Algorithm~\ref{alg:dataset} summarizes the construction pipeline after filtering. The veterinarian-designed schema in Table~\ref{tab:schema} defines the ground-truth label space, and the 416-image gold seed set provides the reference annotations for teacher selection. Doubao-Seed-2.0-Pro obtains the highest field-wise exact-match agreement and is used to annotate the filtered image pool. We retain reports that are valid under the schema, canonicalize their values, and remove missing images and duplicate reports. This process yields 28,696 unique image--report pairs. We allocate 26,087 pairs to the structured training set and 2,609 pairs to the held-out test set. The test set is excluded from all training phases.

The natural-language targets used in Phase 1 and the compact targets used in Phase 3 are derived from selected reports in the structured training set. They reuse the corresponding images and do not increase the number of unique images. Counting each image with its phase-specific target, CatellectBench provides approximately 38K training instances and 2K held-out test instances. Table~\ref{tab:phases} summarizes the target format and role of each set, while Table~\ref{tab:dataset_composition} reports the base split and annotation coverage.

\begin{table*}[t]
\centering
\small
\renewcommand{\arraystretch}{1.18}
\setlength{\tabcolsep}{5pt}
\begin{tabular*}{\textwidth}{@{\extracolsep{\fill}}p{0.10\textwidth}p{0.13\textwidth}p{0.67\textwidth}@{}}
\toprule
\textbf{Set} & \textbf{Data Size} & \textbf{Target Format and Role} \\
\midrule
Phase 1 & 1K &
Controlled natural-language behavior descriptions derived from structured behavior reports; used to learn feline behavior semantics before structured prediction. \\
\addlinespace[2pt]
Phase 2 & 26K &
Structured behavior reports in JSON format; used to learn schema-constrained prediction with FAW loss. \\
\addlinespace[2pt]
Phase 3 & 10K &
Compact serialization of the same structured behavior fields; used to shorten generation while retaining deterministic restoration to JSON. \\
\addlinespace[2pt]
Held-out test & 2K &
Structured behavior reports in JSON format; excluded from all training phases and used to evaluate cat presence, counting, and fine-grained behavior prediction. \\
\bottomrule
\end{tabular*}
\caption{CatellectBench instances used in the three post-training phases and held-out testing. Phase 1 and Phase 3 reuse images selected from the Phase 2 structured training set.}
\label{tab:phases}
\end{table*}

\begin{table*}[t]
\centering
\small
\renewcommand{\arraystretch}{1.12}
\begin{minipage}[t]{0.465\textwidth}
\centering
\begin{tabular*}{\linewidth}{@{\extracolsep{\fill}}lrrr@{}}
\toprule
\textbf{Base Split} & \textbf{Size} & \textbf{Cat} & \textbf{w/o Cat} \\
\midrule
Structured training & 26K & 96.4\% & 6\% \\
Held-out test & 2K & 80\% & 20\% \\
\bottomrule
\end{tabular*}
\vspace{2pt}
\centerline{(a) Dataset split statistics}
\end{minipage}
\hfill
\begin{minipage}[t]{0.465\textwidth}
\centering
\setlength{\tabcolsep}{2pt}
\begin{tabular*}{\linewidth}{@{\extracolsep{\fill}}lrrrrrr@{}}
\toprule
\textbf{Coverage Band} & \textbf{0} & \textbf{1--5} & \textbf{6--9} & \textbf{10--13} & \textbf{14--17} & \textbf{18--22} \\
\midrule
Training ratio (\%) & 6.44 & 0.76 & 3.22 & 36.26 & 25.84 & 27.49 \\
\bottomrule
\end{tabular*}
\vspace{2pt}
\centerline{(b) Field\mbox{-}richness score distribution}
\end{minipage}
\caption{CatellectBench dataset composition. (a) Sample size and proportions for images with visible cats and images without cats (w/o Cat). (b) Percentage of training samples grouped by field\mbox{-}richness score. Larger scores indicate that more observable feline\mbox{-}behavior schema fields are populated in the JSON annotation.}

\label{tab:dataset_composition}
\end{table*}

The structured training set contains approximately 96.4\% cat images and 6\% no-cat images. The held-out test set uses an 80:20 ratio to place greater emphasis on cat-presence generalization while retaining positive examples for count and fine-grained behavior evaluation.

\begin{figure*}[t]
\centering
\includegraphics[width=1.0\textwidth]{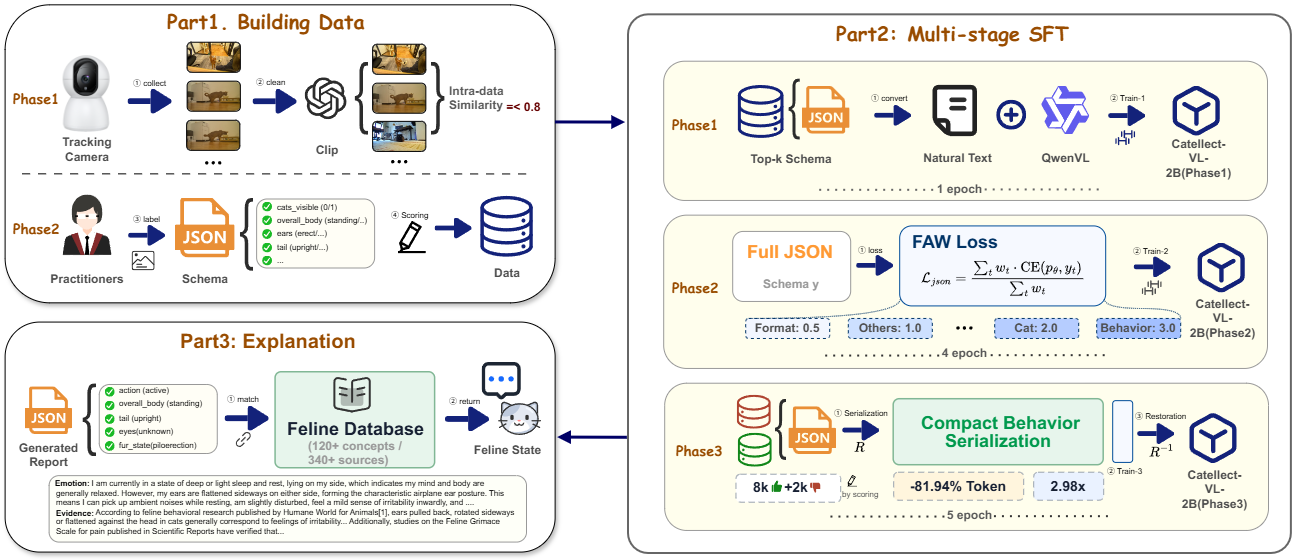}
\caption{Overview of the Catellect-VL-2B training and inference pipeline. CatellectBench is constructed through image filtering, teacher selection, and schema validation. Catellect-VL-2B is then post-trained through natural-language behavior warmup, Field-Aware Weighted SFT, and compact behavior serialization. At inference time, compact predictions are restored to structured behavior fields and matched against the feline behavior knowledge base for evidence-supported interpretation.}
\Description{An overview diagram of CatellectBench construction, three-phase Catellect-VL-2B post-training, compact decoding, and knowledge-grounded inference.}
\label{fig:overview}
\end{figure*}

\subsection{Feline Behavior Knowledge}

For downstream interpretation, we build a feline behavior knowledge base that takes the structured behavior fields predicted by Catellect-VL-2B as input. Table~\ref{tab:schema} lists these fields and their label spaces.

The knowledge base contains 3,151 field-rule entries organized around 120 behavior concepts across six categories: feeding and drinking, sleeping and resting, grooming and hygiene, communication and emotion, interaction and play, and health-related anomalies. These entries are formed from valid value combinations of seven structured behavior fields: action, overall body posture, ears, tail, eyelid, mouth, and fur state. Each field-rule entry defines an exact mapping from one field combination to an evidence-supported interpretation. Therefore, the interpretation layer performs exact field-rule matching rather than open-ended semantic retrieval: after the structured behavior report is validated and unknown or empty values are canonicalized, its fields are matched against the knowledge base.

The knowledge base returns a feline emotion or intent concept, supporting evidence, a risk level, and evidence citations. The rightmost panel of Figure~\ref{fig:target_formats} shows an example produced by matching the predicted structured behavior fields against the knowledge base.

The knowledge base is constructed through a Codex (GPT-5.6-Sol) retrieval-and-distillation workflow. To avoid treating an LLM as an ungrounded authority, the workflow is constrained in three ways. First, the same PhD-trained licensed veterinarian recommends the source scope, including feline behavior guidelines, veterinary behavior reviews, pain-assessment studies, and university or professional veterinary resources \citep{ellis2013aafpisfm, ellis2018felineemotions, evangelista2019feline}. Second, the Codex agent searches only within this source scope and records source metadata for each fetched document. Third, each distilled entry must keep the field-rule format: feline emotion or intent is attached to observable structured behavior fields, supporting evidence is written against those fields, and evidence citations are stored with the entry. The final knowledge base uses 45 curated source profiles and 340 fetched source records.

\section{Multi-Phase Post-Training}
\label{sec:post-training}
\subsection{Overview}

Catellect-VL-2B is post-trained from Qwen3-VL-2B to predict structured behavior fields from a cat image. A separate knowledge-grounded interpretation module matches the predicted fields against the feline behavior knowledge base. This separation keeps visual prediction inside the model and allows the knowledge base to be updated without retraining it. Algorithm~\ref{alg:Catellect} summarizes post-training and inference.

\subsection{Three-Phase Post-Training}

Catellect-VL-2B is trained in three phases. Phase 1 learns feline behavior semantics from controlled natural-language targets. Phase 2 learns the JSON schema and its structured behavior fields. Phase 3 learns a shorter output that can be restored to the same JSON schema. Table~\ref{tab:phases} summarizes the data used in each phase.

\textbf{Phase 1: Natural-Language Behavior Warmup.}
The Phase 2 annotations are structured behavior reports, so training on them directly couples behavior learning with JSON-format learning. We first use 1,200 reports with a visible cat, known action and body-posture labels, and at least one observed ear, tail, or face cue. These criteria focus the warmup on images with explicit behavioral evidence.

Let $\mathcal{I}_{\mathrm{nat}}$ denote the selected indices in the structured training set $\mathcal{D}_{\mathrm{json}}=\{(x^{(n)},y^{(n)})\}_{n=1}^{N}$. For each selected report, GPT-5.6-Sol uses a fixed rewriting prompt $p_{\mathrm{nat}}$ to produce a controlled natural-language behavior description:
\begin{equation}
r^{(i)} = V_{\mathrm{sol}}(y^{(i)};p_{\mathrm{nat}}),
\quad i\in\mathcal{I}_{\mathrm{nat}},
\end{equation}
where $V_{\mathrm{sol}}(\cdot)$ denotes the rewriting function and $r^{(i)}$ is the resulting target. The prompt preserves the structured behavior fields and prohibits new visual evidence. Figure~\ref{fig:target_formats} shows the natural-language target together with the JSON and compact targets for the same image. The warmup set is:
\begin{equation}
\mathcal{D}_{\mathrm{nat}}
=
\{(x^{(i)}, r^{(i)})\mid i\in\mathcal{I}_{\mathrm{nat}}\}.
\end{equation}

We train on $\mathcal{D}_{\mathrm{nat}}$ for one epoch and compute loss only on assistant tokens $\mathcal{A}(r)$. The warmup objective is:
\begin{equation}
\begin{split}
\mathcal{L}_{\mathrm{warm}}(\theta)
=&
-\frac{1}{|\mathcal{D}_{\mathrm{nat}}|}
\sum_{(x,r)\in\mathcal{D}_{\mathrm{nat}}}
\frac{1}{|\mathcal{A}(r)|} \\
&\times
\sum_{t\in\mathcal{A}(r)}
\log p_{\theta}(r_t \mid x,r_{<t}).
\end{split}
\label{eq:loss_warm}
\end{equation}

This phase connects visual features with feline behavior semantics before schema-constrained prediction is introduced.

\begin{table}[t]
\centering
\small
\renewcommand{\arraystretch}{1.12}
\setlength{\tabcolsep}{3pt}
\begin{tabular}{p{0.13\columnwidth} w{c}{0.13\columnwidth} p{0.70\columnwidth}}
\toprule
\textbf{Category} & \textbf{Weight} & \textbf{Fields} \\
\midrule
Format & 0.5 & Braces, commas, quotation marks, fixed schema keys \\
Context & 1.0 & Lighting, location, anchors, nearby beings, interactions, abnormalities, occlusion fields \\
Cat & 2.0 & cats\_visible and cat-count related values \\
Behavior & 3.0 & action, overall\_body, ears, tail, face, fur\_state \\
\bottomrule
\end{tabular}
\caption{Token categories and loss weights in field-aware weighed SFT. Each supervised assistant token is categorized by the schema field or format span it belongs to.}
\label{tab:loss_weights}
\end{table}

\begin{table*}[t]
\centering
\small
\renewcommand{\arraystretch}{1.12}
\setlength{\tabcolsep}{5pt}
\begin{tabular*}{\textwidth}{@{\extracolsep{\fill}}p{0.20\textwidth}p{0.16\textwidth}p{0.54\textwidth}@{}}
\toprule
\textbf{Rule} & \textbf{Symbol} & \textbf{Meaning} \\
\midrule
Top-level fields & $|$ &
Separates the fixed top-level fields, including cats\_visible, lighting, other\_beings, cats, interactions, environment\_anomalies, and summary. \\
Cat instances & \# &
Separates multiple cat instances. \\
Cat fields & comma &
Separates the fixed slots inside each cat instance. \\
Nested lists & semicolon &
Separates repeated values inside nested list fields. \\
Nested objects & tilde &
Separates fields inside a compact nested object. \\
Interaction records & caret, plus &
Caret separates fields inside one interaction record, and plus separates multiple interaction records. \\
Empty values & @n, @e &
@n represents a null value or an empty list, and @e represents an empty string. \\
Escaping & backslash &
Escapes separator symbols when they appear inside text values. \\
\bottomrule
\end{tabular*}
\caption{Compact serialization rules used in Phase 3. The compact format removes fixed JSON keys and repeated punctuation, while preserving a deterministic path back to the same schema fields.}
\label{tab:compact_rules}
\end{table*}

\textbf{Phase 2: Field-Aware Weighed SFT.}
We then train Catellect-VL-2B on the 26K structured training set for four epochs. Each target is a valid structured behavior report under the feline behavior schema. Standard JSON SFT assigns the same loss weight to every assistant token, including braces, commas, quotation marks, and fixed schema keys. These tokens are necessary for parseability but do not carry the same visual information as field values. We therefore assign each supervised assistant token to a format, context, cat-presence, or behavior category. Table~\ref{tab:loss_weights} defines the categories and their weights.

Let $\mathcal{D}_{\mathrm{json}}=\{(x^{(n)}, y^{(n)})\}_{n=1}^{N}$ be the structured training set, where $x^{(n)}$ is an image and $y^{(n)}$ is its structured behavior report. For each assistant token $y_t$ with $t\in\mathcal{A}(y)$, where $\mathcal{A}(y)$ denotes the assistant-token positions, let $c_t$ be its token category. The weight $w_t$ is assigned from $c_t$ as:
\begin{equation}
w_t =
\begin{cases}
0.5, & c_t=\mathrm{Format},\\
1.0, & c_t=\mathrm{Context},\\
2.0, & c_t=\mathrm{Cat},\\
3.0, & c_t=\mathrm{Behavior}.
\end{cases}
\end{equation}

For one training example $(x,y)$, the FAW loss is:
\begin{equation}
\ell_{\mathrm{json}}(\theta;x,y)
=
\frac{
\sum_{t\in\mathcal{A}(y)}
w_t
\left[-\log p_{\theta}(y_t \mid x,y_{<t})\right]
}{
\sum_{t\in\mathcal{A}(y)} w_t
}.
\end{equation}

The full Phase 2 objective is:
\begin{equation}
\mathcal{L}_{\mathrm{json}}(\theta)
=
\frac{1}{|\mathcal{D}_{\mathrm{json}}|}
\sum_{(x,y)\in\mathcal{D}_{\mathrm{json}}}
\ell_{\mathrm{json}}(\theta;x,y).
\end{equation}

This phase keeps the full JSON interface while shifting the learning signal toward cat presence and behavior-critical visual fields.

\begin{algorithm}[t]
\caption{Post-Training and Inference}
\label{alg:Catellect}
\begin{algorithmic}[1]
\REQUIRE Structured training set $\mathcal{D}_{\mathrm{json}}$, base model $f_{\theta_0}$, rewriting function $V_{\mathrm{sol}}$, serializer $R$, restore function $R^{-1}$, schema paths $\Pi$, knowledge base $\mathcal{K}$, input image $x_*$
\ENSURE Structured behavior report $\hat{y}$ and evidence-supported interpretation $z$

\STATE \textbf{Post-training}
\STATE Select 1,200 reports with explicit action, posture, and body-part cues from $\mathcal{D}_{\mathrm{json}}$.
\STATE Build $\mathcal{D}_{\mathrm{nat}}$ by rewriting the selected reports as controlled natural-language descriptions with $V_{\mathrm{sol}}$.
\STATE Train $f_{\theta_0}$ on $\mathcal{D}_{\mathrm{nat}}$ for one epoch to obtain $f_{\theta_1}$.
\STATE Train $f_{\theta_1}$ on $\mathcal{D}_{\mathrm{json}}$ for four epochs with Field-Aware Weighted SFT to obtain $f_{\theta_2}$.
\STATE Build compact targets $R(y)$ and keep only samples satisfying $R^{-1}(R(y))\equiv_{\Pi}y$.
\STATE Train $f_{\theta_2}$ on $\mathcal{D}_{\mathrm{cmp}}$ to obtain Catellect-VL-2B $f_{\theta_3}$.

\STATE \textbf{Knowledge-grounded inference}
\STATE Generate a compact behavior string $\hat{s}\leftarrow f_{\theta_3}(x_*)$.
\STATE Restore a structured behavior report $\hat{y}\leftarrow R^{-1}(\hat{s})$.
\STATE Match $\hat{y}$ against $\mathcal{K}$ with exact field-rule matching.
\STATE Produce evidence-supported behavior interpretation $z$.
\RETURN $\hat{y}, z$
\end{algorithmic}
\end{algorithm}

\textbf{Phase 3: Compact Behavior Serialization.}
Structured behavior reports are useful for supervision and auditing, but their JSON representation is long to generate. Phase 3 keeps the same structured behavior fields and replaces the JSON surface form with a compact fixed-order string. Starting from the Phase 2 model, we serialize each report into a shorter target and train the model to generate this representation. A deterministic restore function then maps the compact string back to the JSON format used for evaluation and knowledge grounding.

Let $\Pi=(\pi_1,\ldots,\pi_m)$ be the ordered list of schema paths kept in the compact format, where $\pi_i$ denotes one schema field path. Given a structured behavior report $y$, the serializer $R$ produces a compact string:
\begin{equation}
s = R(y) =
\operatorname{Join}_{i=1}^{m}\psi_i(y[\pi_i]),
\end{equation}
where $s$ is the compact target, $y[\pi_i]$ is the value at path $\pi_i$, and $\psi_i(\cdot)$ maps the value to its slot-specific compact form. Table~\ref{tab:compact_rules} summarizes the main serialization rules. Since the field order is fixed, the compact string does not need to generate field names, braces, repeated punctuation, or empty containers that can be recovered from the schema.

We define a deterministic restore function $R^{-1}$ and keep a compact target only if it passes the round-trip check:
\begin{equation}
R^{-1}(R(y)) \equiv_{\Pi} y,
\end{equation}
where $\equiv_{\Pi}$ means that all fields indexed by $\Pi$ match after restoration. The compact training set is:
\begin{equation}
\begin{aligned}
\mathcal{D}_{\mathrm{cmp}}=
\{(x,R(y))\mid &(x,y)\in\mathcal{D}_{\mathrm{src}},\\
& R^{-1}(R(y))\equiv_{\Pi}y\}.
\end{aligned}
\end{equation}
where $\mathcal{D}_{\mathrm{src}}$ is the selected source set for compact training.

Compact separators carry structural information for restoration, so we give them a larger loss weight. For each assistant token $s_t$, the compact token weight is:
\begin{equation}
u_t =
\begin{cases}
2.0, & s_t \text{ contains a compact separator},\\
1.0, & \text{otherwise}.
\end{cases}
\end{equation}

For one compact training example $(x,s)$, the loss is:
\begin{equation}
\ell_{\mathrm{cmp}}(\theta;x,s)
=
\frac{
\sum_{t\in\mathcal{A}(s)}
u_t
\left[-\log p_{\theta}(s_t \mid x,s_{<t})\right]
}{
\sum_{t\in\mathcal{A}(s)} u_t
}.
\end{equation}

The Phase 3 objective is:
\begin{equation}
\mathcal{L}_{\mathrm{cmp}}(\theta)
=
\frac{1}{|\mathcal{D}_{\mathrm{cmp}}|}
\sum_{(x,s)\in\mathcal{D}_{\mathrm{cmp}}}
\ell_{\mathrm{cmp}}(\theta;x,s).
\end{equation}

We construct the 10K compact training set from Phase 2 reports that cover a broad range of populated structured behavior fields. Each compact target must pass the round-trip check before training. The shorter output therefore preserves the fields required for evaluation and knowledge grounding.

\subsection{Knowledge-Grounded Interpretation}

After the three-phase post-training, Catellect-VL-2B provides a parseable structured behavior report rather than a user-facing explanation. The knowledge-grounded interpretation module takes this report as input and maps its observable fields to behavior concepts in the feline behavior knowledge base $\mathcal{K}$. Given a restored report $\hat{y}$, the module extracts behavior-relevant fields such as action, overall body posture, ears, tail, face, fur state, abnormalities, interactions, and environment anomalies. It then applies exact field-rule matching against $\mathcal{K}$, so each matched behavior concept is tied to explicit visual fields instead of a free-form caption.

The final output is an evidence-supported interpretation $z$, including a likely behavior interpretation, supporting visible fields, and a conservative risk cue when needed. This separation keeps the system inspectable and allows the knowledge base to be updated without retraining Catellect-VL-2B.

\begin{figure*}[t]
    \centering

    \begin{minipage}[t]{0.32\textwidth}
        \centering
        \includegraphics[width=\linewidth]{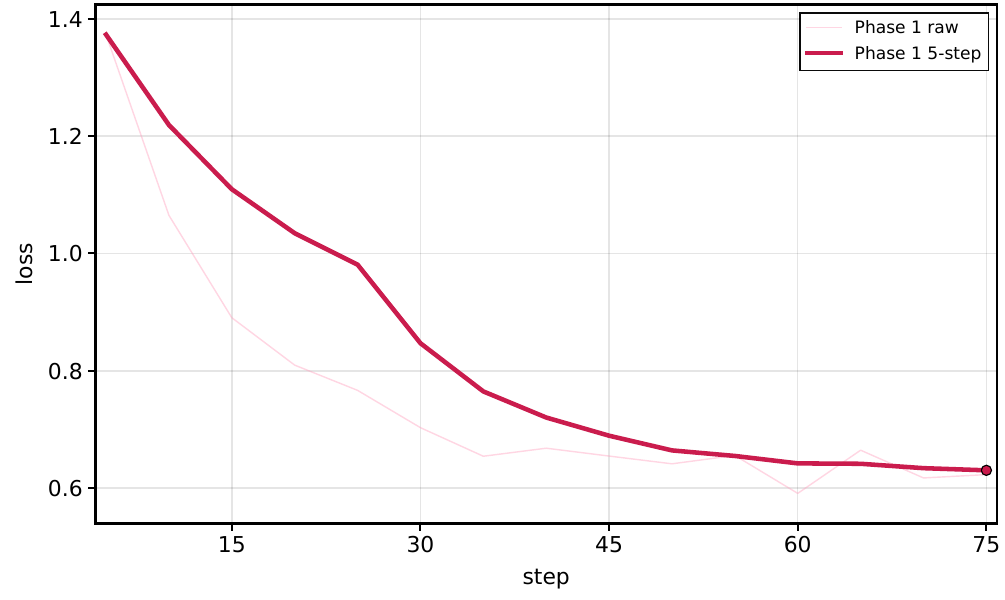}
        \vspace{-2mm}
        \centerline{(a) Phase 1 warmup loss}
    \end{minipage}
    \hfill
    \begin{minipage}[t]{0.32\textwidth}
        \centering
        \includegraphics[width=\linewidth]{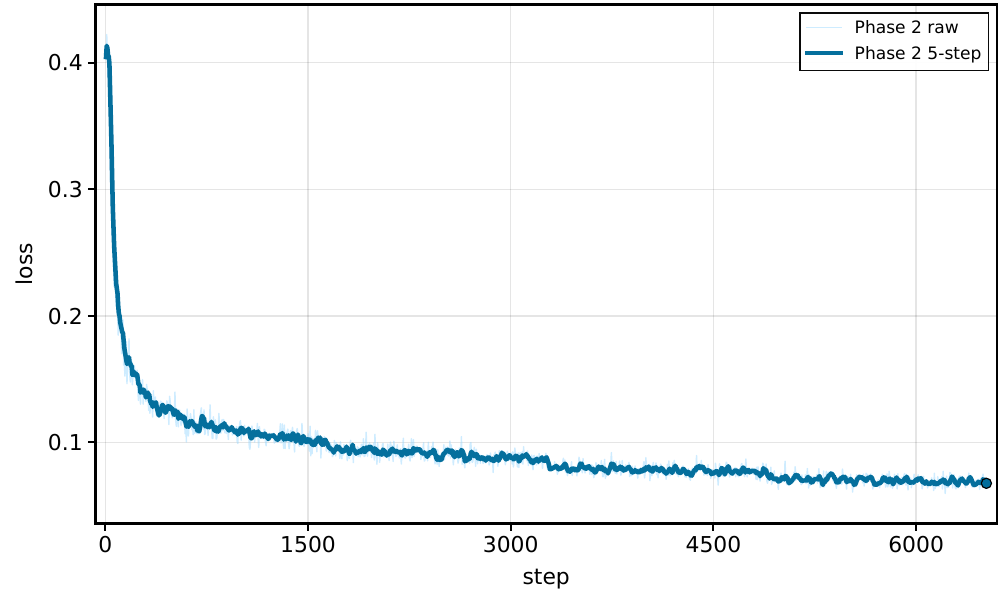}
        \vspace{-2mm}
        \centerline{(b) Phase 2 weighted SFT loss}
    \end{minipage}
    \hfill
    \begin{minipage}[t]{0.32\textwidth}
        \centering
        \includegraphics[width=\linewidth]{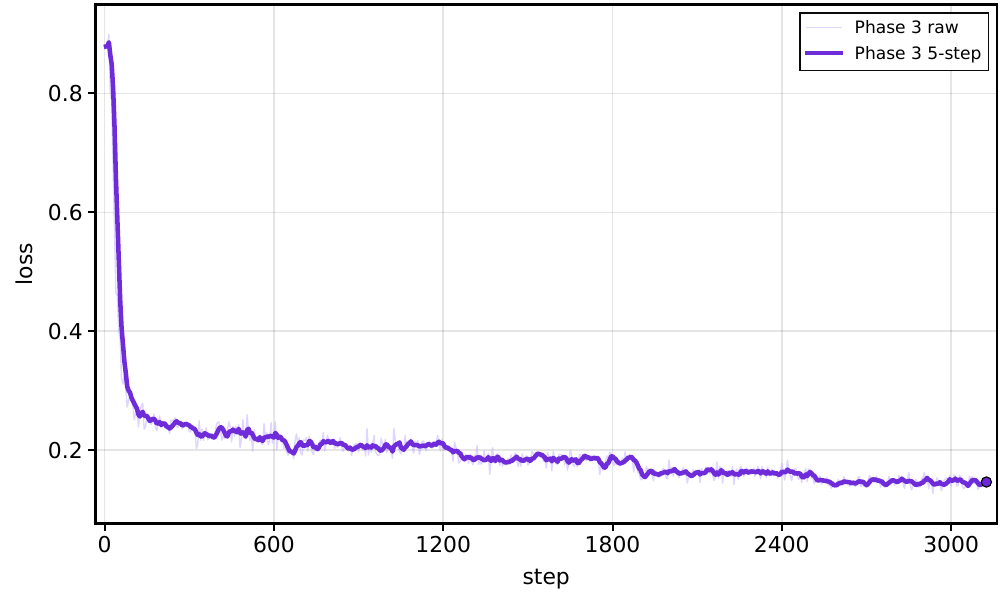}
        \vspace{-2mm}
        \centerline{(c) Phase 3 compact SFT loss}
    \end{minipage}

    \caption{Training loss curves of the three-phase training pipeline. (a) performs warmup training, (b) conducts weighted full-data SFT, and (c) further adapts the model with compact SFT data.}
    \Description{Three line charts showing decreasing training loss during natural-language warmup, weighted JSON supervised fine-tuning, and compact supervised fine-tuning.}
    \label{fig:three_phase_loss}
\end{figure*}

\section{Experiments}

\subsection{Setup and Metrics}

We conduct evaluations covering structured perception, knowledge\mbox{-}base retrieval, compact serialization, inference serving speed, and layer\mbox{-}pruning diagnostics.
All main\mbox{-}table evaluations (Table~\ref{tab:vlm_baselines}, Table~\ref{tab:perception_ablation}, Table~\ref{tab:compact}) run under SDPA\mbox{-}enabled evaluation.
JSON\mbox{-}format evaluation uses greedy decoding with \texttt{do\_sample=False}, \texttt{max\_new\_tokens=2048}, and batch size 2.
Compact\mbox{-}serialization evaluation uses \texttt{max\_new\_tokens=192} and batch size 1.
For training, we adopt LoRA \citep{hu2022lora} with rank 8 and alpha 32 for the JSON parser branch, an effective batch size of 16, maximum sequence length 2560, and image maximum pixels set to 262\,144.

Perception metrics include parse rate, cat\mbox{-}recognition accuracy, cat\mbox{-}count accuracy, strict behavior exact match, behavior\mbox{-}field macro accuracy, and environment\mbox{-}interaction macro accuracy.
For cat\mbox{-}presence binary classification, we further compute true positive (TP), true negative (TN), false positive (FP), and false negative (FN) for confusion\mbox{-}matrix analysis.
Behavior\mbox{-}field macro accuracy averages scores over action, overall body, ears, tail, face, and fur state.
Within tables, \texttt{overall\_body} is abbreviated as Body and \texttt{fur\_state} as Fur.
Environment\mbox{-}interaction macro accuracy averages lighting, location, vertical position, nearby beings, interaction prop type, and interaction anchor type.

\subsection{Open\mbox{-}Weight and Hosted VLM Baselines}
Table~\ref{tab:vlm_baselines} compares Catellect\mbox{-}VL\mbox{-}2B against hosted and open\mbox{-}weight vision\mbox{-}language models under the identical structured\mbox{-}output setting, including the original untuned Qwen3\mbox{-}VL\mbox{-}2B backbone.
Hosted closed\mbox{-}source models are included only as reference points and are excluded from best\mbox{-}result highlighting.
All open\mbox{-}weight baselines are evaluated on the CatellectBench held\mbox{-}out test set, sharing identical input images, schema prompt, JSON parser, and field\mbox{-}level evaluation metrics.
This comparison addresses a natural reviewer question: can off\mbox{-}the\mbox{-}general\mbox{-}purpose VLMs simply be prompted to produce our target JSON schema?
Results show generic VLMs can frequently generate syntactically valid JSON. Even so, they deliver weak performance on cat\mbox{-}specific behavior fields, especially fine\mbox{-}grained body\mbox{-}part cues for ears, tail, and face.

\begin{table*}[t]
\centering
\scriptsize
\setlength{\tabcolsep}{3pt}
\renewcommand{\arraystretch}{1.00}
\resizebox{\textwidth}{!}{%
\begin{tabular}{llrrrrrrrrrr}
\toprule
\textbf{Model} & \textbf{Access} & \textbf{Params} & \textbf{Cat} & \textbf{Count} & \textbf{Action} & \textbf{Body} & \textbf{Ears} & \textbf{Tail} & \textbf{Face} & \textbf{Fur} & \textbf{Avg} \\
\midrule
\multicolumn{12}{c}{\textit{Hosted VLMs (closed-source)}} \\
Gemini3-Pro & hosted & -- & 96.00 & 93.80 & 74.00 & 74.90 & 87.00 & 68.80 & 57.00 & 95.20 & 80.84 \\
GPT-5.4 & hosted & -- & 96.30 & 96.4.00 & 74.30 & 75.00 & 87.50 & 69.10 & 57.80 & 95.30 & 81.16 \\
Claude-Opus-4.8 & hosted & -- & 95.70 & 93.40 & 73.50 & 74.20 & 86.30 & 68.50 & 56.50 & 95.10 & 80.40 \\
\midrule
\multicolumn{12}{c}{\textit{Open-weight VLMs}} \\
Llama-3-LLaVA-NeXT & open & 8B & 80.29 & 77.40 & 5.34 & 40.21 & 0.00 & 0.00 & 0.00 & 91.46 & 36.84 \\
Qwen3-VL-2B & open & 2B & 84.62 & 82.45 & 27.63 & 43.97 & 42.80 & 29.57 & 25.68 & \runner{93.00} & 53.72 \\
MiniCPM-V 4.5 & open & 8B & \textbf{92.07} & 87.26 & \runner{71.68} & 34.77 & 49.10 & 24.37 & 63.80 & 84.59 & 63.46 \\
Qwen3.5-VL-9B & open & 9B & 90.62 & 88.22 & 53.51 & 46.86 & \runner{84.87} & 58.30 & \runner{85.98} & 92.25 & 75.08 \\
\rowcolor{black!8} \textbf{Catellect-VL-2B} & open & 2B & \runner{91.11} & \textbf{88.70} & \textbf{77.65} & \textbf{66.53} & \textbf{92.03} & \textbf{69.72} & \textbf{89.64} & \textbf{93.63} & \textbf{83.63} \\
\rowcolor{black!8} \multicolumn{1}{c}{+Phase3} & open & 2B & \runner{91.11} & \runner{88.30} & 70.53 & \runner{64.80} & 90.61 & \runner{67.67} & 81.65 & 90.25 & \runner{80.62} \\
\bottomrule
\end{tabular}}
\caption{Structured perception benchmark on the CatellectBench held-out test set. Bold marks best performance and underline marks the second-best performance among all open-weight entries, including +Phase3 compact variant; ties share the corresponding highlight. Hosted models are closed-source references and are not highlighted even when their absolute numbers are strong. Avg is computed over Cat, Count, Action, Body, Ears, Tail, Face, and Fur. Catellect-VL-2B denotes the Phase\,2 full JSON parser before compact serialization; +Phase3 is its edge-oriented compact serialization variant achieving a 2.98$\times$ generation speedup at the cost of modest accuracy degradation.}
\label{tab:vlm_baselines}
\end{table*}

Among open\mbox{-}weight models, Qwen3.5\mbox{-}VL\mbox{-}9B obtains 75.08\% overall average accuracy. Our Catellect\mbox{-}VL\mbox{-}2B (Phase\,2) reaches 83.63\% overall average accuracy with only 2B parameters, outperforming Qwen3.5\mbox{-}VL\mbox{-}9B by 8.55\%.
The training loss dynamics across our three\mbox{-}phase post\mbox{-}training pipeline are visualized in Figure~\ref{fig:three_phase_loss}.
Compared with its base Qwen3\mbox{-}VL\mbox{-}2B backbone at 53.72\% average accuracy, post\mbox{-}training brings a large performance gain.
The compact +Phase3 variant retains 80.62\% average accuracy for edge deployment, accepting moderate accuracy loss to obtain faster inference.
Closed\mbox{-}source hosted models achieve competitive overall numbers, yet they cannot be deployed locally on edge hardware.

\subsection{Structured Perception: Training\mbox{-}Strategy Ablations}

\begin{table*}[t]
\centering
\scriptsize
\setlength{\tabcolsep}{3pt}
\renewcommand{\arraystretch}{1.00}
\resizebox{\textwidth}{!}{%
\begin{tabular}{lrrrrrrrrr}
\toprule
\textbf{Method} & \textbf{Cat} & \textbf{Count} & \textbf{Action} & \textbf{Body} & \textbf{Ears} & \textbf{Tail} & \textbf{Face} & \textbf{Fur} & \textbf{Avg} \\
\midrule
\multicolumn{10}{c}{\textit{Baseline}} \\
Qwen3-VL-2B & 84.62 & 82.45 & 27.63 & 43.97 & 42.80 & 29.57 & 25.68 & 93.00 & 53.72 \\
\midrule
\multicolumn{10}{c}{\textit{Single-phase Training}} \\
Vanilla SFT & 80.53 & 77.88 & 74.65 & 63.38 & 88.73 & 64.32 & 87.79 & 93.43 & 78.84 \\
Phase2 (5 epochs) & \textbf{91.59} & \textbf{88.96} & \textbf{78.57} & 61.65 & \runner{91.73} & 66.17 & 86.47 & 91.35 & \runner{82.06} \\
Phase2 (10 epochs) & 89.66 & 87.74 & 72.76 & 61.57 & 91.04 & \textbf{70.15} & 86.19 & 90.67 & 81.22 \\
\midrule
\multicolumn{10}{c}{\textit{Two-Phase Training}} \\
Phase1 + Phase2 & \runner{91.11} & \runner{88.70} & \runner{77.65} & \textbf{66.53} & \textbf{92.03} & \runner{69.72} & \textbf{89.64} & \textbf{93.63} & \textbf{83.63} \\
\bottomrule
\end{tabular}}
\caption{Ablation study of SFT strategies on CatellectBench held-out test set.
Bold and underline mark best and second-best performance.
Qwen3-VL-2B is the zero-shot pre-trained baseline.
Vanilla SFT uses uniform cross-entropy loss, while Phase2 applies our Field-Aware Weighted loss with different training epochs.
Cat / Count denote recognition and counting accuracy; Avg is the average of all evaluated metrics.
Dashes indicate unevaluated metrics.}
\label{tab:perception_ablation}
\end{table*}

Table~\ref{tab:perception_ablation} presents ablation experiments for our training pipeline.
Zero\mbox{-}shot Qwen3\mbox{-}VL\mbox{-}2B achieves reasonable performance for cat detection, yet it struggles on fine\mbox{-}grained feline behavior fields.
Standard vanilla JSON SFT improves behavior prediction results, but it degrades accuracy for cat recognition and cat counting. Our Field\mbox{-}Aware Weighted (FAW) loss addresses this issue: weighting loss across different token categories boosts both cat\mbox{-}related metrics and fine\mbox{-}grained behavior performance. Simply increasing training epochs with FAW loss cannot deliver steady improvements. Phase2 training for 5 epochs yields 82.06\% overall average accuracy; extending training to 10 epochs drops average accuracy to 81.22\%, showing more epochs alone do not bring better results.

The full two\mbox{-}phase recipe combining Phase\,1 natural\mbox{-}language warmup and Phase\,2 FAW\mbox{-}loss JSON SFT obtains the best overall average accuracy of 83.63\%.
This demonstrates that natural\mbox{-}language warmup, paired with our weighted loss objective, forms the most effective training strategy for feline structured behavior understanding. The corresponding cat\mbox{-}versus\mbox{-}no\mbox{-}cat detection confusion matrices in Figure~\ref{fig:confusion_matrices} further verify that our multi\mbox{-}phase training improves both fine\mbox{-}grained behavior metrics and cat detection accuracy for reliable edge deployment.


\subsection{Knowledge\mbox{-}Base Interpretation Pipeline}
Table~\ref{tab:kb} describes our three\mbox{-}component knowledge\mbox{-}driven interpretation workflow.
Our feline\mbox{-}behavior knowledge base stores a large set of behavior\mbox{-}field combination entries.
Once model outputs are parsed into structured behavior fields, we perform direct matching against these stored entries.
Since we use exact field matching rather than open\mbox{-}text retrieval, we do not evaluate this component with a separate natural\mbox{-}language retrieval benchmark in our main results.


\begin{table}[t]
\centering
\small
\setlength{\tabcolsep}{4pt}
\renewcommand{\arraystretch}{1.12}
\begin{tabular}{lll}
\toprule
\textbf{Module} & \textbf{Representation} & \textbf{Role} \\
\midrule
Structured perception &
7 structured behavior fields &
Extracts auditable visual cues from cat images \\

Knowledge index &
3,151 field\mbox{-}combination entries &
Matches behavior fields against feline\mbox{-}behavior supporting evidence \\

Interpretation layer &
Matched evidence and risk cues &
Produces evidence\mbox{-}based explanations for pet care \\
\bottomrule
\end{tabular}
\caption{Knowledge\mbox{-}driven interpretation pipeline of Catellect. Instead of searching from open\mbox{-}ended text captions, the interpretation module queries our manually built feline\mbox{-}behavior knowledge base using model\mbox{-}predicted structured behavior fields.}
\label{tab:kb}
\end{table}

\subsection{Compact Serialization: Ablation on Training\mbox{-}Data Scaling}

\begin{table*}[t]
\centering
\scriptsize
\setlength{\tabcolsep}{3pt}
\renewcommand{\arraystretch}{1.00}
\resizebox{\textwidth}{!}{%
\begin{tabular}{llrrrrrrrrr}
\toprule
\textbf{Method} & \textbf{Data} & \textbf{Cat} & \textbf{Count} & \textbf{Action} & \textbf{Body} & \textbf{Ears} & \textbf{Tail} & \textbf{Face} & \textbf{Fur} & \textbf{Avg} \\
\midrule
Compact 2K & 2K & 82.21 & 79.09 & \textbf{70.92} & 51.42 & 84.04 & 27.66 & \runner{80.50} & \textbf{91.49} & 70.92 \\
Compact 8K & 8K & \textbf{92.87} & \textbf{89.74} & 69.84 & \textbf{65.96} & \textbf{92.46} & \textbf{74.44} & 70.90 & 88.93 & \textbf{80.64} \\
Compact 10K & 10K & \runner{91.11} & \runner{88.30} & \runner{70.53} & \runner{64.80} & \runner{90.61} & \runner{67.67} & \textbf{81.65} & \runner{90.25} & \runner{80.62} \\
\bottomrule
\end{tabular}}
\caption{Phase3 compact serialization ablations on CatellectBench held-out test set, focusing on training data scaling effects. \textbf{Bold} and \underline{underline} mark best and second-best accuracy values. Avg is computed over all evaluated metrics, including Cat, Count, Action, Body, Ears, Tail, Face, and Fur.}
\label{tab:compact}
\end{table*}

Table~\ref{tab:compact} reports ablation results for Phase\mbox{-}3 compact serialization under different training\mbox{-}data sizes.
Training with only 2K compact samples yields poor overall performance. Expanding the compact training set to 8K brings substantial gains across most behavior fields and achieves 80.64\% overall average accuracy. Further expanding training data to 10K yields marginal overall accuracy change, with overall average accuracy dropping slightly to 80.62\%.
We observe that tail\mbox{-}related predictions remain comparatively sensitive under compact representation, even with larger training sets.

Direct mixed training with both JSON replay and compact targets cannot outperform pure compact\mbox{-}target training.
Our Phase\mbox{-}3 compact serialization reduces output length by 81.96.4\%. On Nvidia~A100 it runs at 1.78\,s per image and delivers 2.98× speedup compared with long\mbox{-}JSON generation.
We select the 10K\mbox{-}trained Phase\mbox{-}3 model as our final edge\mbox{-}oriented checkpoint, accepting a small accuracy drop to obtain this critical latency\mbox{-}accuracy trade\mbox{-}off for real\mbox{-}world deployment.

\begin{figure*}[t]
\centering
\setlength{\tabcolsep}{2pt}
\begin{tabular}{cccc}
\includegraphics[width=0.24\textwidth]{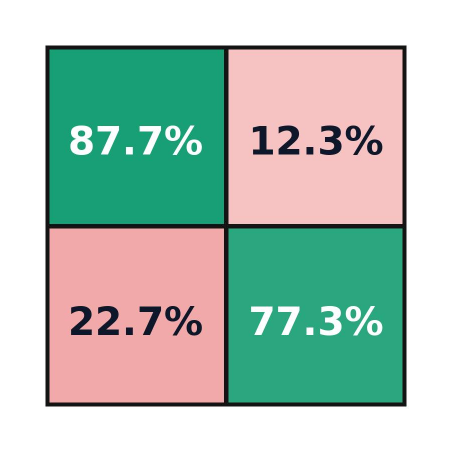} &
\includegraphics[width=0.24\textwidth]{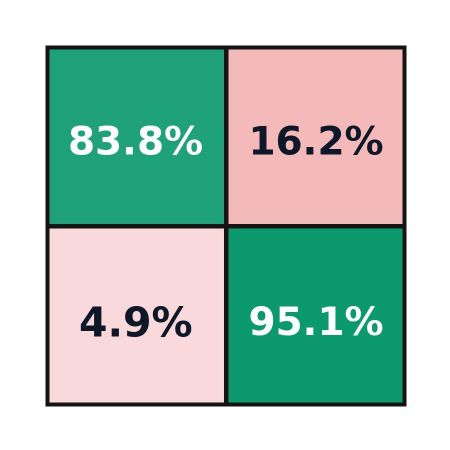} &
\includegraphics[width=0.24\textwidth]{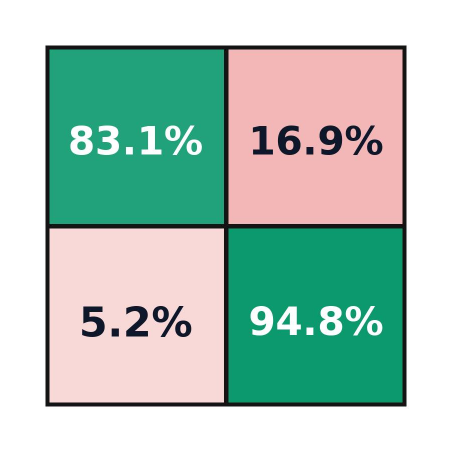} &
\includegraphics[width=0.24\textwidth]{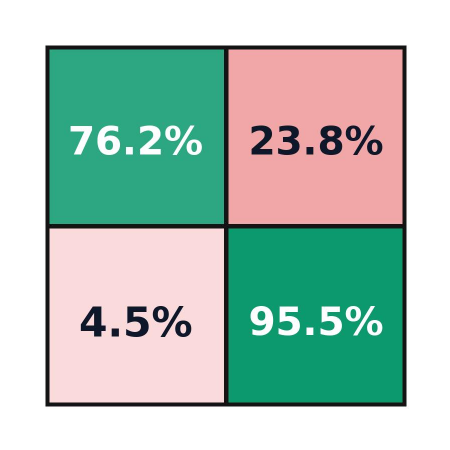} \\
\small (a) Vanilla SFT & \small (b) Phase2  & \small (c) Phase1+Phase2 & \small (d) Phase1+Phase2+Phase3
\end{tabular}
\caption{Confusion matrices for cat presence classification. For each matrix: TP (True Positive): predicted cat present, ground\mbox{-}truth cat present;
TN (True Negative): predicted no cat, ground\mbox{-}truth no cat;
FP (False Positive): predicted cat present, ground\mbox{-}truth no cat;
FN (False Negative): predicted no cat, ground\mbox{-}truth cat present.
Green cells denote high sample counts; pink cells highlight mis\mbox{-}prediction entries.}
\Description{.}
\label{fig:confusion_matrices}
\end{figure*}

\begin{table*}[t]
\centering
\scriptsize
\setlength{\tabcolsep}{3pt}
\renewcommand{\arraystretch}{1.08}
\resizebox{\textwidth}{!}{%
\begin{tabular}{llrrrrl}
\toprule
\textbf{Method} & \textbf{Takeaway} & \textbf{Time (s/img) $\downarrow$} & \textbf{Nvidia A100 $\uparrow$} & \textbf{RK3576 $\uparrow$} & \textbf{Avg} & \textbf{RK3576} \\
\midrule
Eager & Long JSON baseline with full schema decoding & 5.29 & 1.00x & 1.00x & 83.63 & $\surd$ \\
SDPA & Attention-kernel acceleration under the same prompt & 3.80 & 1.39x & -- & 83.63 & $\times$ \\
SGLang + FlashAttention & Server-side continuous batching and serving runtime & 1.46 & 3.62x & -- & -- & $\times$ \\
vLLM + FlashAttention & Server-side paged KV cache with FlashAttention & 1.88 & 2.81x & -- & 83.51 & $\times$ \\
\textbf{Phase-3 compact} & \textbf{Short compact serialization with the same visual model} & \textbf{1.78} & \textbf{2.97x} & \textbf{2.51x} & \textbf{80.62} & $\surd$ \\
H2O & Decode-time KV-cache pruning with extra bookkeeping & 5.59 & 0.95x & -- & 74.46 & $\times$ \\
VisionZip & Prefill-time visual-token pruning before decoding & 3.45 & 1.53x & -- & 41.70 & $\times$ \\
\bottomrule
\end{tabular}}
\caption{Deployment diagnostics normalized to seconds per image. Nvidia A100 speedups are measured relative to the Eager long-JSON baseline. RK3576 speedup is reported only for methods deployable on the Rockchip edge target; server-GPU runtimes and diagnostic pruning baselines are not applicable.}
\label{tab:deployment}
\end{table*}

\subsection{Serving and Pruning Diagnostics}
We evaluate deployment performance across compact\mbox{-}output acceleration, server\mbox{-}side serving acceleration, and token\mbox{-}pruning baselines, with full results listed in Table~\ref{tab:deployment}.
Server\mbox{-}optimized runtimes including SGLang and vLLM deliver strong speedups on Nvidia~A100, but they are incompatible with Rockchip RK3576 edge hardware. The RK3576 provides no exposed KV\mbox{-}cache interfaces and only supports eager decoding, which blocks nearly all cache\mbox{-}centric acceleration techniques. Existing pruning alternatives also show critical limitations for feline behavior understanding: H2O KV\mbox{-}cache pruning incurs non\mbox{-}trivial bookkeeping overhead and yields no net speed gain, whereas VisionZip discards visual tokens and significantly hurts accuracy on fine\mbox{-}grained cat posture and body\mbox{-}part cues.

Our training\mbox{-}driven Phase\mbox{-}3 compact output delivers inference acceleration without custom\mbox{-}backend dependencies and works natively within eager decoding. On RK3576, our compact serialization cuts per\mbox{-}image inference latency from 81.2\,s of the long\mbox{-}JSON baseline down to 32.3\,s while preserving acceptable task accuracy. Such speed improvement meets the latency requirement for edge\mbox{-}side inference and makes real\mbox{-}world on\mbox{-}device deployment feasible.

\section{Conclusion}
Bridging the gap between cats and human caregivers through automated behavior understanding remains an unexplored research direction.
This work presents Catellect\mbox{-}VL\mbox{-}2B, an edge\mbox{-}oriented vision\mbox{-}language model for feline behavior understanding.
Built upon our 40K\mbox{-}sample CatellectBench dataset, our multi\mbox{-}phase post\mbox{-}training pipeline enables the model to predict auditable structured behavior fields and produce compact serializable outputs recoverable to standard JSON format.
Our model outperforms open\mbox{-}weight baselines more than five times larger in parameter size.
Paired with our curated feline\mbox{-}behavior knowledge mapping, it generates evidence\mbox{-}backed behavior interpretations.
Running as an open\mbox{-}weight model on local edge hardware avoids sending user images to remote cloud servers, offering strong privacy guarantees for end users.
When deployed on the RK3576 edge chip, our compact\mbox{-}output variant achieves a 2.51× speedup compared with the full\mbox{-}JSON baseline of the same model size.
Our findings demonstrate that domain\mbox{-}specific supervision together with compact\mbox{-}output serialization boosts feline behavior prediction performance, and lays a practical foundation for real\mbox{-}world edge\mbox{-}device deployment.

\section{Ethical Statement}
Catellect\mbox{-}VL\mbox{-}2B and its knowledge\mbox{-}grounded interpretation module are designed for behavior interpretation and pet\mbox{-}care support only, and are not a substitute for professional veterinary diagnosis.
All health\mbox{-}related outputs are risk\mbox{-}level hints to draw human attention. Severe or persistent animal symptoms must be handled by qualified veterinary professionals.
Even with local edge execution, users should minimize raw image storage and keep full control over data storage, sharing and deletion.

\appendix






\bibliographystyle{ACM-Reference-Format}
\bibliography{sigconf}

\end{document}